# Social Fröhlich condensation: Preserving societal order through sufficiently intensive information pumping

# Andrei Khrennikov

Linnaeus University, International Center for Mathematical Modeling in Physics and Cognitive Sciences, Växjö, SE-351 95, Sweden

**Abstract**

Purpose – This paper aims to present the basic assumptions for creation of social Fröhlich condensate and attract attention of other researchers (both from physics and socio-political science) to the problem of modelling of stability and order preservation in highly energetic society coupled with social energy bath of high temperature.

Design/methodology/approach – The model of social Fröhlich condensation and its analysis are based on the mathematical formalism of quantum thermodynamics and field theory (applied outside of physics).

Findings – The presented quantum-like model provides the consistent operational model of such complex socio-political phenomenon as Fröhlich condensation.

Research limitations/implications – The model of social Fröhlich condensation is heavily based on theory of open quantum systems. Its consistent elaboration needs additional efforts.

Practical implications – Evidence of such phenomenon as social Fröhlich condensation is demonstrated by stability of modern informationally open societies.

Social implications – Approaching the state of Fröhlich condensation is the powerful source of social stability. Understanding its informational structure and origin may help to stabilize the modern society.

Originality/value – Application of the quantum-like model of Frhlich condensation in social and political sciences is really the novel and original approach to mathematical modeling of social stability in society exposed to powerful information radiation from mass-media and internet based sources.

# 1 Introduction

In the dictatorial society, order is preserved through establishing numerous constraints, including restrictions on information delivery and absorption. Since information carries social energy, such information restrictions can be considered as restrictions on highly energetic information flows (censorship, information filtering).[1] The state authorities understand well the social energizing power of information and try to restrict it via the physical control. In spite such restrictions, some individuals are able to approach the state of high social energy and be stationary in such states. The corresponding population shifts can become substantial.

In all dictatorial societies these passional people are the permanent disturbing factor. Physical struggle with those highly excited shifts of society and even elimination active people from social life (prisons, concentration camps, death penalties) led to distraction of economics, science, and literature and arts.

In terms of social temperature, we can say that dictatorial ruling is directed to freezing of society, so to say up to absolute zero. As we known from physics, total isolation and freezing is expansive and demands a lot of energy. To preserve social stability through isolation, the regime should pay double price in social energy: 1) essential part of social energy is used to eliminate passional societal shifts; 2) the energy of passional people is simply destroyed without being used for the needs of society. This social energy misuse is one of the main causes leading dictatorial regimes to collapse. Their collapse, in spite of approaching high homogeneity in social energy structure of population.

What is an alternative to the "freezing control of society" (based on restriction of information flows carrying social energy)? This is the democratic control based on powerful information flows and creation of information reservoir (bath) with high social temperature. How can such a highly energetic stability be modeled mathematically? It happens that the corresponding model is widely used in bioscience and it is known as the model of Fröhlich condensation (Fröhlich, 1968a,b, 1970, 1972, 1977). This phenomenon can be mathematically structured in the framework of quantum-like modeling similarly to the recently developed theory of information thermodynamics (Khrennikov, 2004, 2005, 2010a) and social (information) laser (Khrennikov, 2015b,2016, 2018ab, 2019, 2020a,b, Khrennikov et al., 2018, 2019; Tsarev et. al., 2019; cf. with genuine quantum theory for biological F¨olich condensation, Wu and Austin, 1981, Zhang, Agarwa, and Scully, 2019).

---

[1] Sometimes such restrictions can take an absurd form, e.g., in Soviet Union all taping devices were rigidly controlled, especially before communist celebrations, as November 7.

*Quantum-like models* reflect the features of biological, cognitive, and socio-political processes which naturally match the quantum formalism. In such modeling, it is useful to explore *quantum information theory,* which can be applied not just to the micro-world of genuine quantum systems. Generally, systems processing information in the quantum-like manner need not be quantum physical systems; in particular, they can be macroscopic biological or social systems. Surprisingly, the same mathematical theory can be applied at all biological scales: from proteins, cells and brains to humans and social systems; we can speak about *quantum information biology and sociology* (Asano et al., 2015a). We remark that during the last 10 years quantumlike modelling flowered by attracting the interest of experts in cognition, psychology, decision making, economics and finance, social and political sciences (see Asano et al., 2015b, Bagarello, 2012, Basieva and Khrennikov, 2017, Busemeyer and Bruza, 2012, Busemeyer et al,. 2014, Dubois, 2009, 2014, Dubois and Toffano, 2016, Khrennikov, 2010b, 2015a, Khrennikova, 2016, 2017, Toffano, 2020a,b, for a sample of papers; googling on "quantum-like" gives around 250 000 references). Applications to social and political sciences are not restricted to theory of social laser, also see (Haven and Khrennikov, 2013, Robinson and Haven, 2015, Haven, Khrennikov and Robinson, 2017).

We also remark that above discussion on concentration of population of dictatorial societies at the lowest social energy state can be modeled with the social analog of the phenomenon of Bose-Eintsein condensation. In this paper we restrict the analogy to just this remark, we plan to elaborate this thema in one of further publications.

## 2     Basic components of the model, social analogs of conditions leading to Fröhlich condensation

We suggest to apply the Fröhlich formalism to social energy and systems. We start with establishing correspondence between the components of the Fröhlich model (Fröhlich, 19681,b, 1970, 1972, 1977) and social entities. We use quantum-like formalization of works on social laser (Khrennikov, 2020a,b). Fröhlich by himself did not appeal directly to the quantum formalism. He used the methods of mesascopic physics and thermodynamics in the spirit of early Einstein's works on spontaneous and stimulated emission and absorption of quanta of the electromagnetic field. The quantum reformulation of the Fröhlich model was done in paper (Wu and Austin, 1981) and completed in the recent paper of Zhang, Agarwa, and Scully (2019). For us it is convenient to proceed within similar framework, because the quantum methodology provides the possibility to define the basic social entities formally, as quantum observables, and without to go deeply in social, psychological, cognitive, and even neurophysiological issues.

Although the quantum-like formalization is based on the standard technique of open quantum systems (Ingarden, Kossakowski and Ohya, 1997), Fr¨ohlich condensation is a very delicate phenomenon and its occurrence is constrained by a few conditions of thermodynamical and quantum information nature. We formulate these conditions within discussions of the corresponding components of the model and then summarize them in section 3. On the other hand we are not aimed to present the formal quantum-like derivation in the social framework; this will be done in one of the future works. The aim of the present paper is to discuss the basic social issues related to Fröhlich condensation.

### 2.1    The original biological model

Fröhlich considered in a biosystem (1) oscillating segments of giant dipoles in macromolecules and (2) a heat bath; say protein molecular in a cell filled with solution. The system is open and it is exposed by an external energy pump which couples to the oscillating units. Each unit is involved in processes of a) direct energy absorption from external supply, b) energy exchange with the heat bath, c) redistribution of energy between the levels ($E_i$). Frhlich found conditions on supply, bath and biounits which lead to concentration of all excitations in a biounit at the lowest positive energy mode $E_1$. But, this energy mode can be sufficiently high, i.e., this is not the Bose-Einstein type condensation around zero temperature. In contrast, temperature $T$ of surrounding bath has to be sufficiently high.

## 2.2 Social energy and social atoms

The notion of social energy (*S*-energy) has already been elaborated in very detail in social laser theory (Khrennikov, 2020a,b). In contrast to works of psychologists and sociologists (starting with James, Freud, Jung), we introduce *S*-energy operationally, as an observable on social systems. Mathematically it is described as a quantum obseravble, i.e., by a Hermitian operator acting in the complex Hilbert space of mental states of humans. A human, a discrete indivisible system, is a social analogue of atom - *S*-atom. Any *S*-atom is characterized by its *S*-energy spectrum:

$$E_{0a} < E_{1a} < E_{2a} < ... < E_{Ma}. \tag{1}$$

The ground state mode $E_{0a} \approx 0$ corresponds to the total relaxation of *S*-atom, the state of the passive rest; $E_{1a} > 0$ is the lowest active state mode.

In our previous works on social lasing, we considered mainly two level *S*-atoms (or following the physical lasing technology, 3-4 level *S*-atoms). The Fröhlich formalism handle multilevel systems, i.e., $n$ can be sufficiently large. This is more realist situation; humans are socially complex systems and their energetic behavior is characterized by the multilevel structure.

Mental states are characterized by *S*-energy and additional variables (similar say to photon's polarization or direction) which were called in monograph (Khrennikov, 2020a,b) quasicolor of the mental state, $|\psi\rangle = |E_{ka}\alpha\rangle$, where $E_{ka}$ is the energy level and $\alpha$-quazicolor. In this paper, the supplementary social obseravbles are not considered and only the *S*-energy states ($|E_{ka}\rangle$) are of the interest. Generally *S*-atom's state can be in superposition of these eigenstates:

$$|\psi\rangle = \sum_{j}^{n} c_j |E_{ka}\rangle, \tag{2}$$

where $c_j$ are complex probability amplitudes, $\sum_{j}^{n} |c_j|^2 = 1$. They encode the probabilities to find *S*-atom in the corresponding *S*-energy states

$$p_j = |c_j|^2. \tag{3}$$

We repeat once again that the *S*-energy of the *S*-atom should not be treated as its objective property, this is just a possible output of some special measurement procedure. In the same way, the physical energy of photon is not photon's internal property, this is an outcome of its interaction. We keep to the Copenhagen interpretation and do not assign any objective meaning to superposition (2). Following to Schrödinger, we consider it as an expectation catalog for outcome of measurement.

## 2.3 Information reservoir - social energy bath

One of the basic elements of the model is *the information reservoir.* It is composed of all possible types of information stored in published newspapers, journals, movies, videos, internet based social networks. The reservoir includes the business-information sea, including data on economics, finance, social and political situations, medicine and medical care, science (online lectures and videos, webinars, Wiki). *S*-atoms are involved in information exchange with this reservoir. They emit as well as absorb information excitations in the form of conversations, comments and posts, e.g., in YouTube, Tweeter, or Telegram, different data bases.

Since information excitations carry quanta of *S*-energy, the information reservoir around *S*-atoms is at the same time the *S*-energy bath. And *S*-atoms exchange with it by quanta of *S*-energy. (We generally explore the duality between flows of information and social energy.)

One of the basic conditions for derivation of the Fröhlich condensation regime is the validity of the Planck formula for the average number of excitations corresponding to the concrete energy level. Consider the *S*-energy bath characterized by the spectra:

$$E_{0b} < \ldots < E_{kb} \tag{4}$$

Then the Planck expression has the form:

$$N_{ib} = \frac{1}{e^{E_{ib}/\lambda} - 1}, \tag{5}$$

where the parameter $\lambda$ has the dimension of energy. In physics, $\lambda = KT$, where $T$ is temperature and $K$ is the Boltzmann constant.

This formula can be derived by using the Gibbs methods of virtual ensembles (see Schrödinger, 1989). It is applicable to any type of systems, including social systems. To get the Bose-Einstein statistics, we have to assume that *S*-energy quanta are *indistinguishable* (and exclude the Fermi-Dirac and parastatistics).

Indistinguishability of information quanta is a consequence of very big volume of information stored in the reservoir. *S*-atoms are not able to analyze consciously the information content which delivered permanently via numerous channels coupling it to the reservoir, sms, comments, posts, video,... . The main characteristic of such information exchange is *S*-energy content. Of course, this is not complete contentless-exchange. However, *S*-atom distinguishes communications by just a few parameters forming aforementioned quazicolor $\alpha$. This is clip-thinking, popcorn brain functioning, violation of the laws of Boolean logic (see Khrennikov, 2020a,b) for details.

## 2.4    Quantum information field

Another basic mathematical entity of the quantum-like model is the quantum information field. It formalizes mathematically external supply of *S*-energy into *S*-atoms population. It is generated by massmedia including internet based information sources; this delivery is charaterized by concentration on a few social energy modes. Consider the *S*-energy spectra of external information field:

$$E_{0e} < ... < E_{me.} \qquad (6)$$

The number of modes on which this field in concentrated is essentially less than in the information reservoir, in (4) $k$ is essentially larger than $m$ in (6). These are so to say the basic information flows, they are mainly highly energetic. Thus energy is delivered to the corresponding high energetic modes of *S*-atoms. The essence of Fröhlich condensation is that the high *S*-energy modes of *S*-atoms redistribute energy to lower energy modes and finally to the lowest active mode $E_{1a}$; a part of *S*-energy is emitted into *S*-energy bath (the information reservoir).

The field is the formal operator-valued entity expressed via the operators of creation and annihilation. We describe external *S*-energy supply with quantum information field. Mass-media including the internet based resources generate quanta of information associated with communications, the excitations of the field. Some quanta are absorbed inside the field (the information exchange between information delivery agencies), other quanta are absorbed by *S*-atoms. To behave quantumly, the information field should also satisfy the *indistinguishability* constraint, i.e., its excitations have to be identical up to their energy content. We repeat that we follow the Copenhagen interpretation and indistinguishability is feature of these quanta w.r.t. an observer, *S*-atom. The content indistinguishability regime is approached via *information overload*, i.e., field's intensity should be so high that *S*-atom would not be able to analyze communications carried by the field consciously. *S*-atom distinguish quanta only w.r.t. their energizing content. This model is oversimplified and in a more general model we consider additional information components of communications, quazicolor of the information excitation.

Indistinguishability and information overload go in tact with Fröhlich condensation: for the latter, intensity I of energy supply has to overcome some threshold $I_0$, i.e.

$$I \geq I_0 \tag{7}$$

The expression for threshold $I_0$ depends on energy spectra and a few other model's parameters.

## 2.5 Storage of social energy

The $E_{1a}$-mode is used for *S*-energy storage. In the absence of external energy supply, the population of *S*-atoms approaches the state of thermodynamics equilibrium and the majority of *S*-atoms are concentrated at the ground state $|E_{0a}\rangle$. This is the simple consequence of the Planck formula:

$$n_{ia} = \frac{1}{e^{E_{ia}/\lambda'} - 1}. \tag{8}$$

The term with $E_{0a} \approx 0$ strongly dominates over other terms and $n_{0a} \gg n_{1a} > \ldots > n_{Ma}$. Only for sufficiently high I (see (7)) the tendency of *S*-atoms to relax completely can be overcome and they can be concentrated at the active states with $i > 0$. This possibility to store energy in society is one of the distinguishing features of social Fröhlich condensation.

## 2.6 Social temperature

The formal operational determination of social temperature is based on determination of a class of measurement procedures and calibration. In contrast to physical temperature, such procedures are not well elaborated. Another possibility is to determine social temperature thermodynamically.

For the equilibrium state the parameter $\lambda$ can be determined from the Planck formula (11):

$$\lambda = \frac{E_{ib}}{\ln(1 + 1/N_{ib})}. \tag{9}$$

The crucial feature of the equilibrium state that the right-hand side does not depend on *i*; it is homogeneous w.r.t. the energy spectrum of the reservoir, in our case the information reservoir. This feature can be used as a test for approaching of the information equilibrium state.

Then social temperature can be defined as scaling of $\lambda$ by any parameter *k* with dimension energy/temperature. So,

$$T = \frac{E_{ib}}{k \ln(1 + 1/N_{ib})}. \tag{10}$$

(Even in physics selection $k = K$ is just the subject for the agreement.) Thus we can write the Planck formula in same way as in physics:

$$N_{ib} = \frac{1}{e^{E_{ib}/kT} - 1}. \tag{11}$$

The information reservoir with high social temperature is characterized by the large number of information communications carrying highly *S*-energetic excitations. If social temperature is low, then the majority of information communications carry low *S*-energetic excitations. Now we are ready to formulate social analog of the high temperature regime: $kT \gg E_{1a}$. The information reservoir has to be hot and characterized by the presence of the large number of "exciting communications", news, videos, tweets, comments. In particular, the presence of a variety of hot news is the important condition for creation of social Fröhlich condensate.

One of the determining constants on social bath leading to Fröhlich condensation is the high temperature regime, i.e.,

$$\lambda = kT \gg E_{1a}. \tag{12}$$

So, social temperature of the information reservoir has to be sufficiently high. Fröhlich condensation is impossible in cold social bath. This phenomenon is possible only in the information reservoir which is full of hot news, comments, posts, sms, and conversations. For societal stability, such hot information is crucial. This is the very special feature of this phenomenon: order and stability not via social cooling, but via social heating: *more shock news and information - higher degree of stability.*

## 2.7 Summary for conditions leading to Fröhlich condensation

1. Discreteness of *S*-energy spectra for socio-information systems: *S*-atoms, social bath (information reservoir), information field; see (1), (4), (6).
2. Indistinguishability, up to *S*-energy, of information excitations (more generally up to some additional characteristics - quazicolor).
3. Bose-Einstein statistics of *S*-excitations filling social bath - the Planck formula (11).
4. Sufficiently high intensity of external information supply, (7).
5. High temperature of social bath (information reservoir), (12).
6. Big *S*-energy capacity of social bath.

# 3 Stability and order preservation in highly informationally energized society

Social realization of the phenomenon of Fröhlich condensation provides a mathematical model for order preservation in the informationally open society such as in the democratic states of Europe, USA, Canada, and Australia. In contrast to the dictatorial states such as North Corea, China, Russia, Iran, the open societies are characterized by the absence of censorship and restrictions on information distribution.[2] To some surprise, intensive information supply, high social temperature in combination with the big information reservoir lead to concentration of social energy at the lowest active energy mode $E_{1a}$. This is done via redistribution of *S*-energy between the energy levels of *S*-atoms. The information reservoir (social bath) also absorbs a part of energy. But the main part of energy is stored at the lowest nontrivial mode $E_{1a}$. In this way the open society solves jointly the two problems: a) peaceful elimination of passional part of population; b) sustainable functioning at the energy mode $E_{1a}$. This mode, although not very high, i.e., $E_{1a} << E_{Ma}$, is still essentially high, typically $E_{1a} >> E_{0a}$. The majority or population is full of social energy which is sufficient for active economic and societal life. At the same time the high energy states $E_{ja}, j >> 1$, are not attractors and an $S$-atom with high probability makes transtion from $|E_{ja}>$ to $|E_{1a}>$ (generally via a chain of transition via intermediate states). In spite of the absence of censorship and rather smooth actions of the state repressive apparatus, society escapes mass protests and revolutions, including color revolutions.[3] Of course, small scale protests can happen even inthe open societies, but their have relatively small amplitude and they are not dangerous for stability of the democratic regimes.

---

[2] Of course, we discuss the ideal situation. In reality even "open societies" are not totally open; even in say USA censorship exists, but it has more intelligent forms than say in Russia.

Although stability of the society and active and creative stability can be considered as its strong side, the total elimination of passional modes makes society gray-homogeneous. This is the good place to cite "Screwtape" of Lewis (1994):

"You are to use the word ("democracy") purely as an incantation; if you like, purely for its selling power. It is a name they venerate. And of course it is connected with the political ideal that men should be equally treated. You then make a stealthy transition in their minds from this political ideal to a factual belief that all men are equal. Especially the man you are working on. As a result you can use the word democracy to sanction in his thought the most degrading (and also the least enjoyable) of human feelings. You can get him to practise, not only without shame but with a positive glow of self-approval, conduct which, if undefended by the magic word, would be universally derided. The feeling I mean is of course that which prompts a man to say I am as good as you.

The first and most obvious advantage is that you thus induce him to enthrone at the centre of his life a good, solid, resounding lie. I don't mean merely that his statement is false in fact, that he is no more equal to everyone he meets in kindness, honesty, and good sense than in height or waist measurement. I mean that he does not believe it himself. No man who says I m as good as you believes it. He would not say it if he did. The St. Bernard never says it to the toy dog, nor the scholar to the dunce, nor the employable to the bum, nor the pretty woman to the plain. The claim to equality, outside the strictly political field, is made only by those who feel themselves to be in some way inferior. What it expresses is precisely the itching, smarting, writhing awareness of an inferiority which the patient refuses to accept."

---

[3] This apparatus is the basic part of the state machines of even open societies. And a part of $S$-atoms in high energy states is controlled and often isolated in so to say unnatural way, i.e., without the Fröhlich condensation process. Drugs also play the important role in struggle with such highly energetic individuals. However, in any way these are just small fractions of population. The social thermodynamics makes the main job automatically.

We repeat once again that indistinguishability is the fundamental property of social systems leading to quantum statistics of $S$-energy distribution and the possibility to use quantum information societal control, including Fröhlich condensation. This is the price for stability of open society.

# 4 Conclusion

Under assumptions 1-6 (section 2.7) we can apply the quantum-like model of social Fr̈ohlich condensation. This coherent condensation of population at the lowest active mode of $S$-energy explains stability of modern informationally open societies. This stability and order preservation are based on natural self-elimination of passional individuals carrying too high $S$-energy; the individuals who in principle can destroy social order and generate various instabilities. This self-regulation is based on $S$-energy redistribution between states of $S$-atoms and active $S$-energy exchange with the information reservoir and external information field. The crucial condition of creation of social Fröhlich condensate is the loss of individuality of humans as well as the loss of the ability for detailed analysis of information based on the rules of classical Boolean logic. The latter is a consequence of information overload generated by external information supply. To create the stable social Fröhlich condensate, this supply should be very intensive - over some threshold depending on model's parameters.